\documentclass[letterpaper]{article} 
\usepackage{aaai23}  
\usepackage{times}  
\usepackage{helvet}  
\usepackage{courier}  
\usepackage[hyphens]{url}  
\usepackage{graphicx} 
\usepackage{natbib}  
\usepackage{caption} 
\usepackage{algorithm}
\usepackage{algorithmic}
\usepackage{booktabs}       

\usepackage{newfloat}
\usepackage{listings}
\DeclareCaptionStyle{ruled}{labelfont=normalfont,labelsep=colon,strut=off} 
\floatstyle{ruled}
\newfloat{listing}{tb}{lst}{}
\floatname{listing}{Listing}
\title{Towards Voice Reconstruction from EEG during Imagined Speech}
\author{
    Young-Eun Lee\textsuperscript{\rm 1}\equalcontrib,
    Seo-Hyun Lee\textsuperscript{\rm 1}\equalcontrib,
    Sang-Ho Kim\textsuperscript{\rm 2},
    Seong-Whan Lee\textsuperscript{\rm 2}\thanks{Corresponding author}
   } 

\affiliations{

    \textsuperscript{\rm 1} Department of Brain and Cognitive Engineering, Korea University, Seoul, Republic of Korea\\
    \textsuperscript{\rm 2} Department of Artificial Intelligence, Korea University, Seoul, Republic of Korea\\
    \{ye\_lee, seohyunlee, sh\_k, sw.lee\}@korea.ac.kr

}

\usepackage{bibentry}

\begin{document}

\maketitle

\begin{abstract}
Translating imagined speech from human brain activity into voice is a challenging and absorbing research issue that can provide new means of human communication via brain signals. Endeavors toward reconstructing speech from brain activity have shown their potential using invasive measures of spoken speech data, however, have faced challenges in reconstructing imagined speech. In this paper, we propose NeuroTalk, which converts non-invasive brain signals of imagined speech into the user's own voice. Our model was trained with spoken speech EEG which was generalized to adapt to the domain of imagined speech, thus allowing natural correspondence between the imagined speech and the voice as a ground truth. In our framework, automatic speech recognition decoder contributed to decomposing the phonemes of generated speech, thereby displaying the potential of voice reconstruction from unseen words. Our results imply the potential of speech synthesis from human EEG signals, not only from spoken speech but also from the brain signals of imagined speech.
\end{abstract}





\section{Introduction}






 

Brain signals contain various information related to human action or imagery, making them valuable materials for understanding human intentions. Brain-computer interface (BCI) is a technology of analyzing user's brain activity to derive external commands to control the environment through brain signals, therefore, can benefit paralyzed or locked-in patients \cite{chaudhary2016brain}. 
Brain-to-speech (BTS) is a novel research stream in the field of BCI, which aims to directly synthesize audible speech from brain signals \cite{lee2022toward, lee2020neural}. While current studies of decoding speech from human brain signals mainly focus on using spoken speech brain signals measured with invasive methods \cite{anumanchipalli2019speech,angrick2019speech,herff2019generating}, reconstructing imagined speech using non-invasive modalities is a fascinating issue that can convert user's imagined speech into real voice. However, due to the fundamental constraint of the imagined speech lacking the ground truth (GT) voice, it is challenging to synthesize user's own voice from imagined speech brain signals.


Since reconstructing speech from brain signals of spoken speech has shown its potential \cite{anumanchipalli2019speech,angrick2019speech,herff2019generating}, we anticipate that there must be a relevant brain activation that may encode significant features of the speech. Imagined speech is known to resemble the neural activation path of spoken speech, which is mainly located on the ventral sensorimotor cortex (vSMC)\cite{watanabe2020synchronization, si2021imagined, cooney2018neurolinguistics, lee2019eeg}. If imagined speech has similar features to spoken speech, it may be possible to link the spoken speech brain signals, spoken speech audio, and imagined speech brain signals. Furthermore, if we could train and infer phonemes from imagined speech, several unseen words composed of already trained phonemes can also be reconstructed from the trained word sets. 

In this study, we proposed NeuroTalk framework that can correlate imagined speech electroencephalography (EEG) with spoken speech EEG and their corresponding audio, to reconstruct voice from imagined speech.
The imagined utterances were decoded from EEG signals to reconstruct voice at a word level. Moreover, we estimated the possibility of reconstructing unseen words using the pre-trained model trained with only few words, to potentially expand the degree of freedom using the model trained with minimal words including various phonemes. Based on our results, we aim to find the potential of speech reconstruction from imagined speech brain signals to the user's own voice. The main contributions are as follows:

\subsection*{Main Contribution}
\label{main_contribution}
\begin{itemize}

\item We propose a generative model based on multi-receptive residual modules with recurrent neural networks that can extract frequency characteristics and sequential information from neural signals, to generate speech from non-invasive brain signals. 

\item The fundamental constraint of the imagined speech-based BTS system lacking the ground truth voice have been addressed with the domain adaptation method to link the imagined speech EEG, spoken speech EEG, and the spoken speech audio. 

\item Unseen words were able to be reconstructed from the pre-trained model by using character-level loss to adapt various phonemes. This implies that the model could learn the phoneme level information from the brain signal, which displays the potential of robust speech generation by training only several words or phrases. 

\end{itemize}


\section{Background}

\subsection{Speech-related Paradigms}
Speech-related paradigms mainly used in the BTS studies can be largely divided into three categories: spoken speech, mimed speech, and imagined speech \cite{schultz2017biosignal}. While spoken speech indicates the natural speech that accompanies vocal output and movement of the articulators, mimed speech does not produce vocal output but accompanies the movement of the mouth and tongue as if speaking out loud \cite{schultz2017biosignal, cooney2018neurolinguistics}. Imagined speech is the mode of internally imagining speech, accompanying both the imagery of the mouth movement and the vocal sound, without producing actual movement or voice \cite{cooney2018neurolinguistics}.

\subsection{Invasive Approach}

Invasive measurements involve surgical process of implementation inside the skull to capture brain activation directly from the cortex. Therefore, medical risks and difficulties exist to be applied for healthy users \cite{wang2021open}. However, due to the high signal-to-noise ratio (SNR), many previous studies focused primarily on synthesizing speech from invasive brain signals. Studies using electrocorticography \cite{akbari2019towards,anumanchipalli2019speech,angrick2019speech,herff2019generating} and attempts to decode speech from deeper brain structures using stereotactic electroencephalography depth electrodes \cite{angrick2021speech,angrick2022towards,angrick2021real, herff2020potential} have reported the possibility of speech reconstruction using spoken speech data. 


\subsection{Non-invasive Approach}

\subsubsection{Electroencephalography}
EEG is the most widely used non-invasive modality for practical use, since it does not involve any surgical process and are relatively easy to access \cite{krishna2020speech}. However, non-invasive measures have relatively low SNR and artifact problems compared to the invasive modalities, which makes it hard to extract user's intention from brain signals \cite{graimann2009brain}.

\subsubsection{Spoken speech based BTS}
Speech reconstruction from spoken speech or mimed speech brain signals, kinematic or EMG data have shown potential \cite{gaddy2020digital,gaddy2021improved,gonzalez2017direct}. However, a spoken speech-based BTS system cannot be the final solution for the essential goal of BCI, since it is not a silent communication (if the user can speak out, there's no need to reconstruct speech from brain signals), and it cannot be used for patients who cannot speak or move.


\subsubsection{Decoding imagined speech}
Current technologies of decoding imagined speech from EEG have shown promising results in terms of classification problems \cite{nguyen2017inferring, saha2019speak, wang2013analysis, lee2020neural, krishna2020speech}. Previous works about imagined speech mostly targeted classification tasks, or text decoding from EEG signals \cite{makin2020machine}. However, it is challenging to expand the number of classes for the classification scenario \cite{lee2020neural}. Also, to provide an intuitive system that can generate voice from the brain signals, speech reconstruction from the imagined speech is crucial.


\subsubsection{Imagined speech based BTS}
The fundamental constraint of speech reconstruction from EEG of imagined speech is the inferior SNR, and the absence of vocal ground truth corresponding to the brain signals. Therefore, speech synthesis from imagined speech with non-invasive measures has so far not led to convincing results \cite{proix2022imagined}. Attempts to reconstruct speech from invasive data during whispered and imagined speech have existed, however, yet have been reported relatively inferior performance with even invasive measures \cite{angrick2021real}. Speech synthesis from imagined speech may be the key to open a new era of human communication from current voice or text-based to brain-based communication. Also, this may be a technology that can help patients who are unable to speak or those who might lose their voice in the future.

\section{Method}

\begin{figure*}[t]
\centering
    \includegraphics[width=0.9\textwidth]{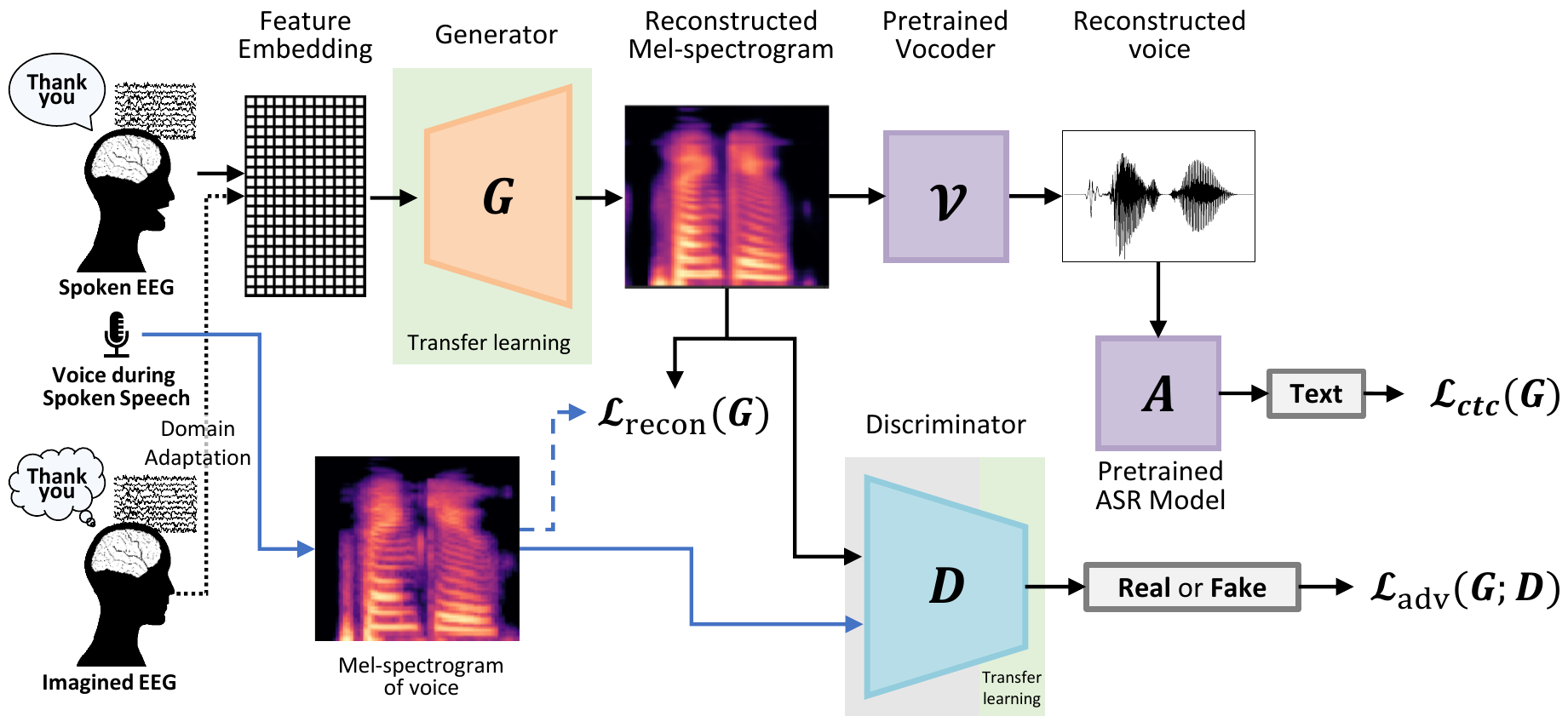}
    \caption{Overall frameworks in this study. Imagined speech EEG were given as the input to reconstruct corresponding audio of the imagined word or phrase with the user's own voice. $G$ refers generator, which generate mel-spectrogram from embedding vector. $D$ refers discriminator, which distinguish the validity of input. On the bottom part, the two model, pretrained vocoder $V$ and a pretrained ASR model $A$, generate text from mel-spectrogram.}
    \label{fig1}
\end{figure*}

\begin{figure*}[t]
\centering
    \includegraphics[width=0.8\textwidth]{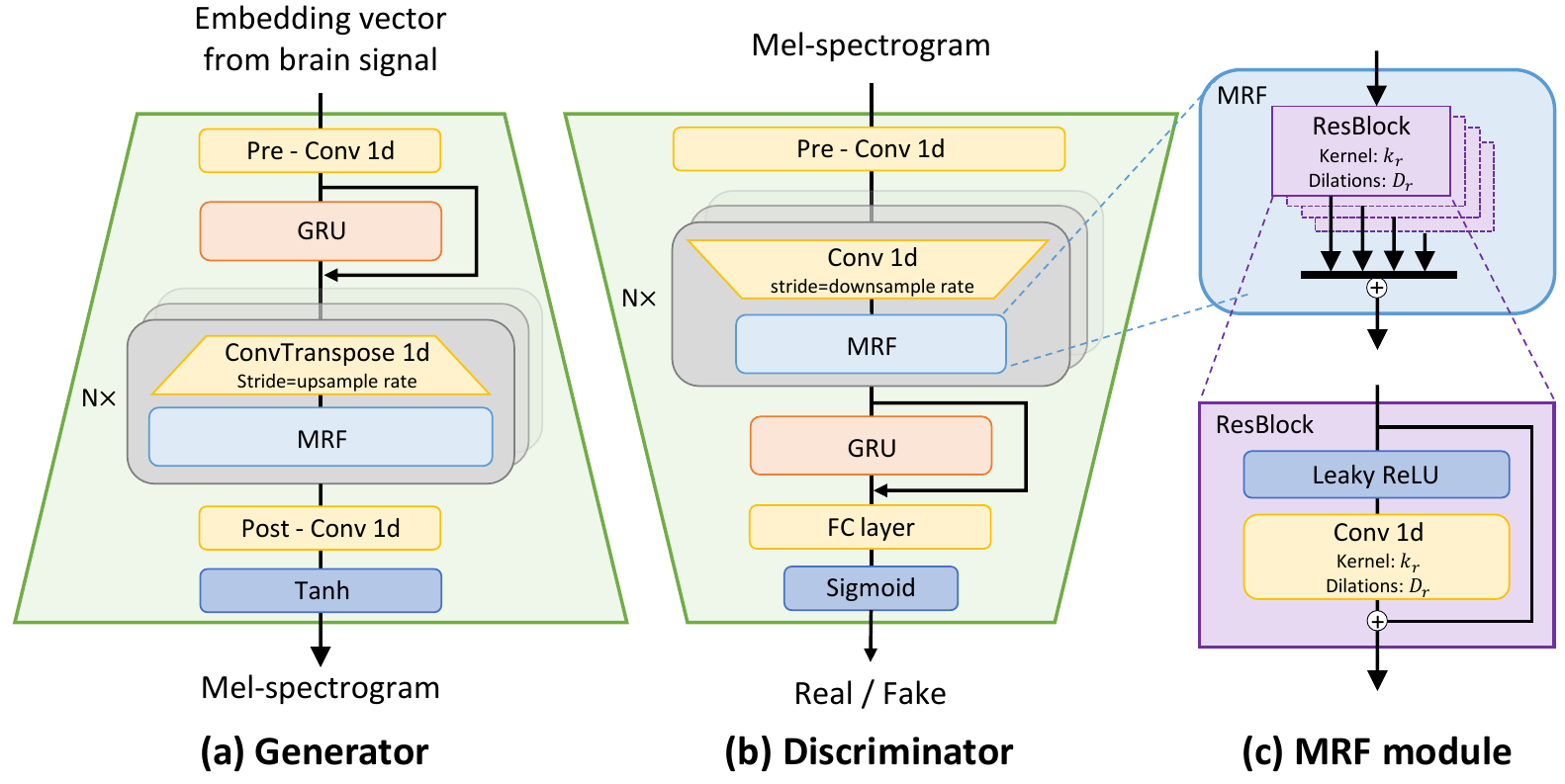}
    \caption{The architecture details in (a) generator, (b) discriminator, and (c) MRF module. The MRF modules in both generator and discriminator were repeated three times in our experiment. $k_r$ indicates the kernel size of residual block and $D_r$ indicates the dilation rates of the residual block.} 
    \label{fig2}
\end{figure*}

In this section, we describe the model frameworks used in this paper, including generator, discriminator, vocoder and automatic speech recognition (ASR), as well as losses including reconstruction loss, generative adversarial network (GAN) loss, and connectionist temporal classification (CTC), as shown in Figure~\ref{fig1}.
The collected brain signals of spoken speech and imagined speech are represented as feature embeddings to extract the optimal features from brain signals. 
The generator applying GAN \cite{goodfellow2014generative} reconstructs a mel-spectrogram to match the target voice during spoken speech. The reconstruction loss for the generator is determined as the difference between the reconstructed mel-spectrogram from the EEG signals and the ground truth mel-spectrogram during spoken speech.
The discriminator classifies the validity of whether the input samples of mel-spectrogram are real or fake, and calculates an adversarial loss for the generator and discriminator.
ASR model is a speech-to-text model, which can represent the speech as a contextual sequence of discrete units \cite{baevski2020wav2vec}. The pretrained vocoder converts the mel-spectrogram to a reconstructed voice, which is then transformed into characters by the pretrained ASR model. The pre-trained ASR model transforms the voice into text, and calculates the CTC loss for the generator.

Since the voices were not recorded during the imagined speech, voices during the spoken speech were used as the ground truth. To match the EEG to the voice of spoken speech, dynamic time warping (DTW) was applied between the reconstructed mel-spectrogram from EEG and the mel-spectrogram of voice during spoken speech. Furthermore, domain adaptation (DA) was conducted to transfer the architecture of spoken speech to that of imagined speech.

\subsection{Architectures}
\subsubsection{Embedding vector}
It is known that spatial, temporal, and spectral information are all important for speech-related brain signals, and vector-based brain embedding features can represent the contextual meaning in brain signals \cite{goldstein2022shared, lee2020neural}. The embedding vector was generated using common spatial pattern (CSP) to maximize spatial patterns and log-variance to extract temporal oscillation patterns. CSP finds the optimal spatial filters using covariance matrices \cite{devlaminck2011multisubject}, and helps to decode the brain signals related to speech \cite{lee2020neural,nguyen2017inferring}. 

To reduce the difference between the data distribution of spoken EEG and imagined EEG, CSP filters were shared with both EEG signals. CSP filters were trained with imagined EEG, which contains pure brain signals only, rather than spoken EEG which may contains some noise. By sharing the CSP filters, spoken EEG domain was adapted to the subspace of imagined EEG.

The CSP filters were trained with eight CSP features and sixteen segments without overlap using training dataset. Each trial of EEG signals has a size of time point $\times$ channels (5000 $\times$ 64). After applying CSP, the embedding vector, transformed from EEG signals, has 104 features $\times$ 16 time segments, where the features consist of 13 classes $\times$ 8 CSP features.

\subsubsection{Generator}
The main architecture of the proposed generator consists of gated recurrent unit (GRU) \cite{cho2014learning} to capture the sequence information and several residual blocks to capture the temporal and spatial information with preventing vanishing gradient issue. Figure~\ref{fig2}a describe the generator in detail. The input of the generator is given as the embedding vector of EEG signals and the output is generated as mel-spectrogram. The embedding vector goes through pre-convolution layer consisting 1d convolution and concatenates the features from bi-directional GRU extract the sequence features. To match the output size of the mel-spectrogram, 1d convoltion layer was applied. After that, the generator upsamples it using transposed convolution with stride of two or three, and multi-receptive field fusion (MRF) module, sum of outputs of multiple residual blocks with different kernel size, follows.

\subsubsection{Discriminator}
The discriminator is similarly composed in the opposite direction to the generator, described in Figure~\ref{fig2}b. The input of the discriminator is the mel-spectrogram and the output is the validity of real/fake voice. Moreoever, the discriminator was trained with their class only using mel-spectrogram from voice. The input goes through pre-convolution layer consisting 1d convolution. And then, the upsampling layer using transposed convolution and MRF module are conducted. After that, the bi-directional GRU extracts the sequence features, and the validity is estimated with the classifier.

\subsubsection{Vocoder and ASR}
Vocoder and ASR model are used to clarify the reconstructed voice from the brain signal by translating to text.
To adjust our framework for a real-time BTS system, we applied a pretrained HiFi-GAN \cite{kong2020hifi} which is a high-quality vocoder with fast inference speed. The same architecture and hyperparameters were applied with the pretrained model `Universal ver.1', trained with Universal dataset.

The ASR is composed of pretrained HuBERT~\cite{hsu2021hubert} with a large configuration, which is a self-supervised learning model of speech representations trained with Libri-Light dataset and fine-tuned with the LibriSpeech dataset.

\subsection{Training Loss Term} 
This section describes losses for training, including reconstruction loss, GAN loss, and CTC loss. The generator uses reconstruction loss $L_{rec}$, adversarial loss $L_{adv}$, and CTC loss $L_{ctc}$, while the discriminator uses adversarial loss $L_{adv}$.
\begin{equation}
L(G) = \lambda_{g1}L_{rec}(G) + \lambda_{g2}L_{adv}(D;G) + \lambda_{g3}L_{ctc}(G)
\end{equation}
\begin{equation}
L(D) = \lambda_{d}L_{adv}(D;G) 
\end{equation}
, where loss coefficients are referred to $\lambda_{g1-3}$ for the generator and $\lambda_{d}$ for the discriminator.

\subsubsection{Reconstruction loss}
To reinforce the guideline for reconstructing the mel-spectrogram of target, reconstruction loss was applied. Reconstruction loss have been verified in many studies \cite{kong2020hifi, isola2017image}, which can help improve the efficiency of generator and the fidelity of reconstructed data. Since imagined speech has no reference speech to compare the reconstructed performance, spoken speech audio collected at the same sequence of imagined speech was used as the target audio to compute reconstruction loss. DTW was applied to match the alignment of EEG during spoken/imagined speech and the target spoken voice.
\begin{equation}
L_{rec}(G) = E_{s}[(G(s)-x)^2]
\end{equation}
, where $s$ refers to the input of the generator such as an embedding vector from EEG signals, and $x$ refers to the input of the discriminator such as a mel-spectrogram.

\subsubsection{GAN loss} 
To reconstruct the mel-spectrogram to follow the real one, adversarial GAN loss $L_{adv}$ was conducted on the generator $G$ and discriminator $D$ as follows.
\begin{equation}
    L_{adv}(D;G) = E_{(x,s)}[log(1-D(x)) + log(D(G(s)))]
\end{equation}
\begin{equation}
    L_{adv}(G;D) = E_{s}[log(1-D(G(s)))]
\end{equation}
, where $x$ refers the input of discriminator such as mel-spectrogram and $s$ refers the input of generator such as embedding vector from EEG signals.

\subsubsection{CTC loss} 
CTC loss is a common metric of the performance for automatic speech recognition systems \cite{graves2006connectionist}. CTC loss $L_{ctc}$ allows to train the model using sequential data without the alignment information. CTC loss was primarily given to guide the prediction of character and phonemes, to enhance the performance of unseen classes.

\subsection{Domain Adaptation}
The DA strategy was employed to resolve the fundamental constraint of speech reconstruction from imagined speech. Since imagined speech does not accompany the movement of the articulators, it is relatively reliable in terms of movement artifacts accompanied by the mouth movement and the vibration. However, since the ground truth audio for imagined speech does not exist, we designed an adaptation framework that adapts the domain of imagined speech from spoken speech, in order to exploit the natural correspondence of imagined EEG and the voice of spoken speech. 
The DA process was performed in two steps; 1) sharing the covariance matrix between imagined EEG and spoken EEG by applying the CSP filter of imagined speech and 2) applying transfer learning for the generator and discriminator from the trained model of spoken EEG.

\subsubsection{Sharing subspace}
The CSP weights, trained with a training set (60\%) of imagined EEG, were shared to generate embedding vectors. Sharing the CSP filters computed from imagined EEG allows the latent space of spoken EEG to be shifted into a comparable feature space of imagined EEG. Unlike most transfer learning approaches that involve applying the weak domain to the well-trained classifier, we elected to the contrary, bringing the spoken speech feature space to that of the imagined speech. In that case, we could achieve a clear pattern more from the brain signal of speech rather than the movement artifacts or vibration artifacts.

\subsubsection{Transfer learning}
The model was trained with a training set of spoken EEG, and then fine-tuned with a training set of imagined EEG at a smaller learning rate than the case of spoken EEG. This was to connect with the voice recordings of spoken speech, which acts as the ground truth of imagined speech. The trained model from spoken EEG can assist training the models of imagined EEG that has insufficient information, therefore, the spoken EEG could guide learning from the weak features of imagined EEG.




\section{Experimental Setup}
\subsection{Dataset}
\subsubsection{Participants}
Six participants volunteered in the study. The study was conducted in accordance with the Declaration of Helsinki, approved by Korea University Institutional Review Board [KUIRB-2019-0143-01]. Informed consent was obtained from all subjects.

\subsubsection{Paradigms}
For the spoken speech session, participants were instructed to naturally pronounce the randomly given thirteen phases, provided as an auditory cue of twelve words/phrases (ambulance, clock, hello, help me, light, pain, stop, thank you, toilet, TV, water, and yes) and a silent phase. Speech data were recorded in a rhythmic manner to avoid any visual or auditory disruptions. The imagined speech data was collected in the exactly same manner as the spoken speech, following the previous study \cite{lee2020neural}. 100 trials of both spoken speech and imagined speech per class were collected for each participant. Therefore, each participant had 1300 trials for the spoken and imagined speech paradigm.

\subsubsection{Recording}
The dataset used in this study consists of scalp EEG recordings of spoken/imagined speech and voice recordings of spoken speech. During the experiment, EEG signals were recorded in the sampling rate of 2500Hz via Brain Vision/Recorder (BrainProduct GmbH, Germany), and the corresponding audio of spoken speech was simultaneously recorded in the sampling rate of 8000Hz. Brain signals were recorded with 64-channel EEG cap with active Ag/AgCl electrode placement following the international 10-10 system.

\subsection{Pre-processing} 
EEG signals were extracted in 2 second intervals for each trial. The data was filtered with a 5th order Butterworth bandpass filter in the high-frequency range of 30--120 Hz which is well-known to contain speech-related information~\cite{lachaux2012high, lee2020neural}. Notch filter was used to remove the line noise at 60 Hz with harmonics of 120 Hz. The electroocoulography (EOG) and electromyography(EMG) of spoken speech were removed using blind source separation referencing from EOG and EMG \cite{gomez2006automatic}. Baseline was corrected by subtracting the average value of 500 ms before each trial. 
Pre-processing procedures were performed in Python and Matlab using OpenBMI Toolbox \cite{leeMH2019eeg}, BBCI Toolbox \cite{krepki2007berlin}, and EEGLAB \cite{Delorme2004EEGLAB}.
For the voice data, we resampled the voice signals to 22050~Hz, and reduced the noise using noisereduce library\cite{Sainburg2019, sainburg2020finding}.

\subsection{Dataset Composition and Training Procedure}
By definition, imagined speech does not have reference voice to train the model. However, spoken speech accompanies vocal output, therefore, both audio and the EEG data of each spoken speech utterance was collected in perfectly time-aligned pair. Since the experimental design of imagined speech and spoken speech was completely identical, the voice recording of the identical sequence of spoken speech for each subject was used as the reference voice for the imagined speech evaluation. Also, due to the lack of the reference voice for the imagined speech brain signals, transfer learning was applied with the model trained on spoken speech EEG and spoken speech audio to imagined speech EEG. 


The dataset was divided in 5-fold into training, validation, and test dataset according to the random selection with random seed. One unseen word, `stop' was separated from the dataset, and wasn't included in the training set. It was chosen to test unseen case, since every phoneme composing the word `stop' was covered with the remaining 11 words used for the training. That is, we trained 11 words/phrases and a silent phase as training dataset, and validated 12 words/phrases and a silent phase in validation and test dataset including the unseen word.

Generator and discriminator were first trained using spoken EEG data with ground truth of voice for each trial. 
As the imagined speech does not have each trial of voice, the voice during spoken speech was used as ground truth. To match the time points between EEG and voice of spoken speech, DTW was applied to the synthesized mel-spectrogram of EEG with mel-spectrogram of voice.
Moreover, the generator and discriminator executed transfer learning from the trained model of spoken EEG to connect naturally between imagined EEG and the voice during spoken speech.

\begin{figure*}[ht]
\centering
    \includegraphics[width=0.9\textwidth]{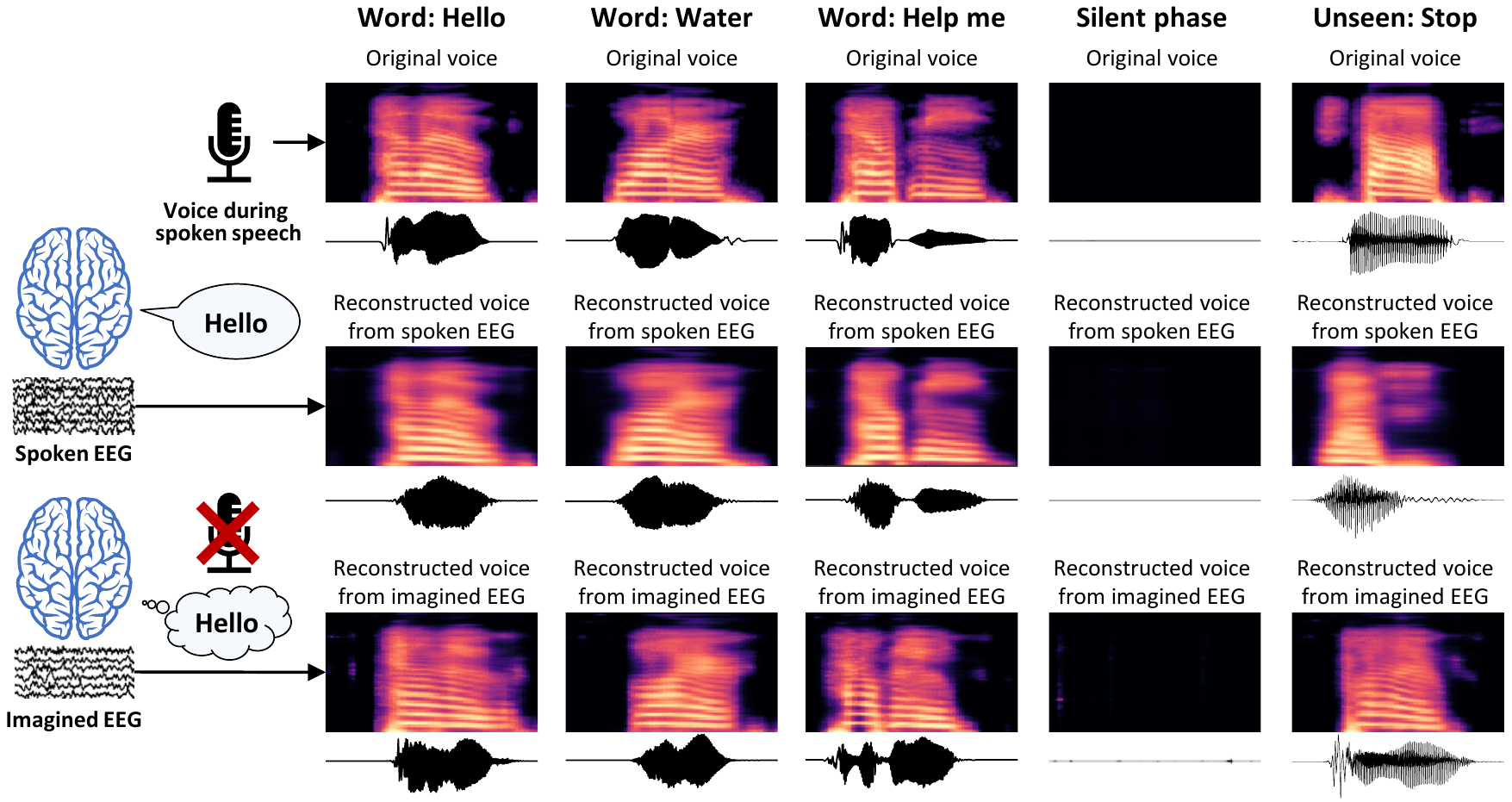}
    \caption{Mel-spectrogram and the audio wave of original voice, reconstructed voice from EEG. Three examples of reconstruction include `Hello', `Water', and `Help me'. Silent phases for both spoken and imagined EEG were successfully decoded. Unseen cases were also reconstructed despite their inferior performance.}
    \label{fig3}
\end{figure*}

\subsection{Model Implementation Details}
The generator had three residual block with the kernel size of 3, 7, and 11, each dilation of 1, 3, and 5, and upsampling rate of 3, 2, and 2 with twice upsample kernel size. The number of initial channel was 1024, the directional GRU dimension was the half of initial channel. The discriminator had the same residual block as generator, but downsampling rate of 3, 3, and 3 with twice kernel size. The number of final channel was 64, the directional GRU dimension was the half of final channel. The mel-spectrogram was managed in sampling rate of 22050~Hz and the STFT and mel function was conducted with nFFT of 1024, window of 1024, hop size of 256, and 80 bands of mel-spectrograms. Initial training was conducted with an initial learning rate of $10^{-4}$, and the fine-tuning was conducted a lower learning rate such as $10^{-5}$ in maximum epoch of 500 and a batch size of 10. We trained the model on a NVIDIA GeForce RTX 3090 GPU. We used AdamW optimizer \cite{loshchilov2017decoupled} with searched parameters of $\beta_1$=0.8, $\beta_2$=0.99, and weight decay $\lambda$=0.01, which was scheduled by 0.999 factor in every epoch. We released the source code and sample data on Github at: https://github.com/youngeun1209/NeuroTalk

\subsection{Evaluation Metrics}
For the evaluation metrics, we used root mean square error (RMSE), character error rate (CER), and a subjective mean opinion score (MOS) test. 
To evaluate the accurate reconstructing performace of the generator, we computed the RMSE between the target and reconstructed mel-spectrogram. 
To evaluate the clarity quantitatively, we conducted CER after going through the ASR model. 
For the subjective evaluation, MOS test was conducted to evaluate the quality of the reconstructed speech. We randomly selected 125 samples of voice from a test dataset. The samples were evaluated by more than 20 raters on a scale of 1-5 with 0.5 point increments. 
We compared EEG dataset with GT and converted GT in form of mel-spectrogram, waveform, and character. 
Moreover, to demonstrate the extension of NeuroTalk, we evaluated the generation performance of unseen word which is composed of phonemes that were contained in the trained word classes.

\section{Results and Discussion}

\subsection{Voice Reconstruction from EEG}
The audio samples are included in the demo page at: https://neurotalk.github.io/demo/neurotalk.html. Figure~\ref{fig3} displays the mel-spectrogram and the audio wave of original voice, reconstructed voice from spoken speech EEG, and reconstructed voice from imagined speech EEG. As shown in the figure, successfully reconstructed cases display similar patterns of mel-spectrogram and the audio waveform. 
Table~\ref{table1} shows the evaluation results of reconstructed voice from brain signal compared with GT. Objective measures of RMSE and CER have shown inferior performance on the case of imagined speech EEG compared to that of the spoken speech EEG. MOS of spoken speech cases was high with no large difference from GT, which means the model can generate natural speech from spoken EEG. Interestingly, the unseen spoken EEG shows higher MOS than imagined EEG though objective evaluation of RMSE and CER indicates inferior, which implies that reconstructing spoken EEG that simultaneously produce voice enables to generate natural voice.

As shown in Figure~\ref{fig3}, test samples for the silent phases were successfully reconstructed with no activation. Silent cases for both spoken and imagined EEG was successfully decoded except for only one case of imagined speech. According to this result, we can infer that our NeuroTalk model accurately learned the silence interval and can detect the precise onset from both spoken speech and imagined speech EEG. Although imagined speech doesn't have ground truth voice, the results show that the proposed NeuroTalk framework effectively adapts the spoken speech based model to the imagined speech EEG to decode user's intention from brain signals and generate voice.

\begin{figure}[t]
\centering
    \includegraphics[width=\columnwidth]{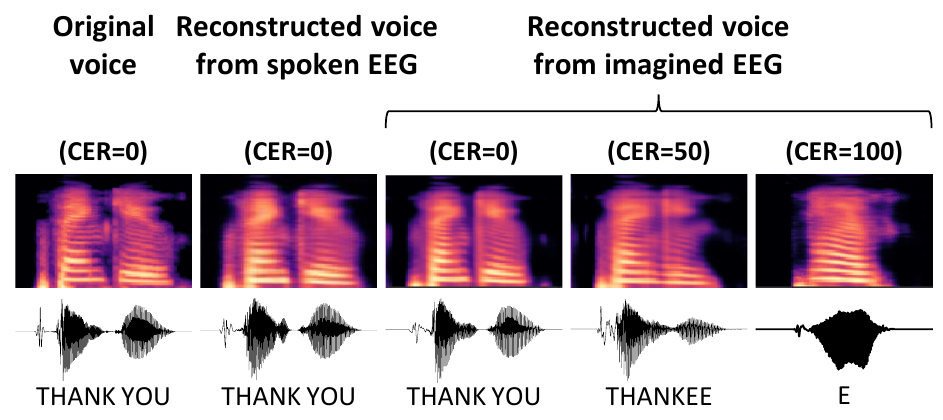}
    \caption{Success and failure cases. Mel-spectrogram and waveform were displayed for original voice and reconstructed voices.}
    \label{fig4}
\end{figure}

There were some instances of failure in the imagined speech case (Figure~\ref{fig4}). The significant difference between the success and failure cases was whether it detects silence intervals. As shown in the Figure~\ref{fig4}, a failure case with CER of 50\% displays few silence interval between 'thank' and 'you'. Moreover, the failure case with CER of 100\% produces only a few words that cannot represent any characters from ground truth.

\subsection{Ablation Study} 
The results of ablation study are demonstrated in Table~\ref{table2}. We performed an ablation study of GRU in generator and discriminator to clarify to add the module in the model, and losses of GAN, reconstruction, and classification to verify the effect of each loss on the performance of generator.
In the objective evaluation, as the performance without reconstruction loss is the worst and then CTC loss, reconstruction loss followed by CTC loss has the greatest impact on the framework. 
Furthermore, the results of all cases were much worse than the baseline, that indicates that all cases are performing their roles, in particular, reconstruction loss have the largest impact on training.
In the subjective evaluation, naturalness shows different tendency to a certain extent, and the results without GRU show the worst with inferior naturalness, which shows that sequential features are important for natural speech synthesis.


\subsection{Voice Reconstruction of Unseen Words}
The unseen and untrained word could be reconstructed from EEG, with much lower CER than chance level, and produce quite high quality of audio with MOS over 3. The gap between the CER of spoken and imagined EEG was relatively small in the unseen case, compared to the trained words. These results imply that while the robust performance of spoken EEG with trained words may be affected by movement artifacts to be overfitted to the trained classes, imagined speech may be relatively effective in capturing phonemes from the trained words to generate unseen word.

Although it still could be further improved, our result demonstrates that the NeuroTalk model has the potential to extend the degree of freedom of decodable words or sentences by training on several word level dataset. We expect that CTC loss could learn the character or phoneme information of words even from brain signals, which contains human intention. Since we trained the model with limited words/phrases, it may be simply classifying the EEG as one of the training classes. However, the model have shown the potential to generate unseen word outside of the training set indicates the possibility of the model to be generalized and expanded to the classes outside of the training set. 

\begin{table}[t]
  \caption{Results of subjective and quantitative tests}
  \label{table1}
  \centering
  \scriptsize
  \begin{tabular}{lllll}
    \toprule
    Model                       & RMSE              & CER (\%)               & MOS \\
    \midrule
    GT                          & -                 & 18.35 ($\pm$11.45)     & 3.67 ($\pm$0.97)    \\
    GT $\rightarrow$ Mel ($\rightarrow$ ASR) & -    & 23.35 ($\pm$10.85)     & 3.68 ($\pm$0.88)  \\
    \midrule
    Spoken EEG                  & 0.166 ($\pm$0.022)  & 40.21 ($\pm$13.49)    & 3.34 ($\pm$0.95)  \\
    Imagined EEG                & 0.175 ($\pm$0.029)  & 68.26 ($\pm$2.47)    & 2.78 ($\pm$1.11)  \\
    Unseen Spoken EEG           & 0.185 ($\pm$0.029)  & 78.89 ($\pm$7.43)    & 2.87 ($\pm$1.12)  \\
    Unseen Imagined EEG         & 0.187 ($\pm$0.026)  & 83.06 ($\pm$14.54)    & 2.57 ($\pm$1.18)  \\
    \bottomrule
  \end{tabular}
\end{table}

\begin{table}[t]
  \caption{Results of ablation study}
  \label{table2}
  \centering
  \scriptsize
  \begin{tabular}{llll}
    \toprule
    Input                       & RMSE              & CER (\%)              & MOS \\
    \midrule
    Baseline                    & 0.175 ($\pm$0.029)  & 68.26 ($\pm$2.47)    & 2.78 ($\pm$1.11)  \\
    \midrule
    w/o GRU                     & 0.185 ($\pm$0.027)  & 76.05 ($\pm$3.25)     & 2.18 ($\pm$1.24)  \\
    w/o GAN loss                & 0.180 ($\pm$0.022)  & 76.05 ($\pm$2.33)    & 2.86 ($\pm$1.21)  \\
    w/o reconstruction loss     & 0.620 ($\pm$0.121)  & 80.16 ($\pm$7.98)     & 2.50 ($\pm$1.25)  \\
    w/o CTC loss                & 0.387 ($\pm$0.069)  & 76.90 ($\pm$0.25)    & 2.52 ($\pm$1.21)  \\
    w/o DA                      & 0.175 ($\pm$0.025)  & 72.30 ($\pm$1.71)     & 2.66 ($\pm$1.24)  \\
    \bottomrule
  \end{tabular}
\end{table}

\subsection{Domain Adaptation} 
DA was performed by sharing the CSP subspaces and transferring the spoken speech-based trained model to imagined speech EEG. As shown in Table~\ref{table2}, the result with DA has shown superior performance compared to the baseline. This implies that spoken speech EEG was useful to train imagined speech EEG, which means the neural substrates of imagined and spoken speech has common features that can be represented in our embedding vector. 
Speech production and articulation is mainly known to be associated with the inferior frontal gyrus, so called Broca's area. Angular gyrus functions to associate various language-related activation from auditory, motor, sensory, and also visual cortex, therefore, not only the left temporal lobe but the whole brain may function in the speech process \cite{watanabe2020synchronization}. Our embedding vector which was generated from the whole channel EEG, may contain both articulatory information and the speech intention.
Therefore, we demonstrates the potential of generating speech by extracting informative speech-related features, which refer to the similarity of spoken speech EEG and imagined speech EEG.

\subsection{Leave-one-out Scenario} 
In order to apply our NeuroTalk system to locked-in patients who can only use imagined speech, we conducted an additional experiment of leave-one-out (LOO) approach to apply to an entirely new data from an unseen person. The model was trained with the spoken EEG of entire subject excluding one subject and was fine-tuned with the imagined EEG of the excluded subject. As a result, comparable performance was obtained, inferior to the baseline but better than without DA. Based on the LOO approach, we have found potential to expand our framework to entirely new person, which could further help people who have lost their own voice. 

\section{Conclusion}
We presented NeuroTalk, which reconstructs user's own voice from EEG during imagined speech. 
DA approach was conducted by sharing feature embedding and training the models of imagined speech EEG, using the trained models of spoken speech EEG. 
Our results demonstrate the feasibility of reconstructing voice from non-invasive brain signals of imagined speech in word-level. Furthermore, unseen word can be generated with several characters although the performance was not high, which means we can expand our study to a larger dataset and to sentence-level speech synthesis in the future. 
We hope our study can contribute to expanding the means of human communication and further benefit patients or disabled people to gain freedom for their communication. 
We look forward to a world where we can communicate without saying anything.




\section{Acknowledgement} 
This work was supported by Institute for Information \& Communications Technology Planning \& Evaluation (IITP) grant funded by the Korea government (MSIT) (No.2021-0-02068, Artificial Intelligence Innovation Hub; No. 2017-0-00451, Development of BCI based Brain and Cognitive Computing Technology for Recognizing User’s Intentions using Deep Learning; No. 2019-0-00079, Artificial Intelligence Graduate School Program(Korea University)).

\bibliography{bib}

\end{document}